\documentstyle[12pt]{article}
\input amssym.def
\input amssym
\def\suma_#1^#2{\setbox0=\hbox{$\scriptstyle{#1}$}
 \setbox2=\hbox{$\displaystyle{\sum}$}
 \setbox4=\hbox{$\scriptstyle{#2}\mathsurround=0pt$}
 \dimen0=.5\wd0 \advance\dimen0 by-.5\wd2
 \ifdim\dimen0>0pt
 \ifdim\dimen0>\wd4 \kern\wd4 \else\kern\dimen0\fi\fi
 \mathop{{\sum}^{#2}}_{\kern-\wd4 #1}}
\newtheorem{theorem}{Theorem}
\newtheorem{definition}{Definition}
\newtheorem{lemma}{Lemma}
\newtheorem{remark}{Remark}

\newenvironment{proof}{\noindent{\it Proof.\/}}{\vskip3mm}
\newenvironment{acknowledgement}
 {{\vskip3mm}{\noindent{\it Acknowledgement.\/}}}{\vskip3mm}

\begin{document}
\title{Deformation Quantization and Nambu Mechanics} \author{Giuseppe
Dito${}^{1}$\thanks{Supported by the European Communities  and the
Japan Society for Promotion of Science.}, Mosh\'e Flato${}^{1,2}$,
Daniel Sternheimer${}^{1,2}$,
Leon Takhtajan${}^{3}$\thanks{NSF grant DMS-95-00557}\\
\\ ${}^1$ Research Institute for Mathematical Sciences\\ Kyoto
University\\ Kitashirakawa, Oiwake-cho\\ Sakyo-ku, Kyoto 606-01
Japan\\ \\ ${}^2$ D\'epartement de Math\'ematiques\\ Universit\'e de
Bourgogne\\ BP 138, F-21004 Dijon Cedex France\\ \\ ${}^3$ Department
of Mathematics\\ State University of New York at Stony Brook\\ Stony
Brook, NY 11794--3651 USA\\} \date{March 8, 1996}
\maketitle
\vspace*{-2mm}\noindent{\bf Abstract:}
Starting from deformation quantization
(star-products), the quantization problem of Nambu Mechanics is
investigated.  After considering some impossibilities and pushing some
analogies with  field quantization, a solution to the quantization
problem is presented  in the novel approach of Zariski quantization of
fields (observables, functions,  in this case polynomials). This
quantization is based on the factorization over ${\Bbb R}$ of polynomials
in several real variables. We quantize the infinite-dimensional
algebra of fields
generated by the polynomials by defining a deformation of
this algebra which is Abelian, associative and distributive. This
procedure is then adapted to derivatives (needed for the Nambu brackets),
which ensures the validity of the Fundamental Identity of Nambu Mechanics
also at the quantum level. Our construction is in fact more  general
than the particular case considered here: it can be utilized for quite
general defining identities and for much more general star-products.
\thispagestyle{empty}
\newpage\setcounter{page}{1}
\section{Introduction}
\subsection{Nambu Mechanics}
Nambu proposed his generalization of Hamiltonian Mechanics \cite{Na}
 by having in mind a generalization of the Hamilton equations of
 motion   which allows the formulation of a statistical mechanics on
 ${\Bbb R}^3$.  He stressed that the only feature of Hamiltonian
 Mechanics that one needs  to retain  for that purpose, is the
 validity of Liouville theorem.  In that spirit, he considered the
 following equation of motion:
\begin{equation}\label{a}
\frac{d\vec{r}}{dt}=\nabla g(\vec{r}) \wedge \nabla h(\vec{r})\; ,
 \quad \vec{r}=(x,y,z)\in {\Bbb R}^3 ,
\end{equation}
where $x$, $y$, $z$ are the dynamical variables and $g$, $h$ are two
functions of $\vec{r}$. Then Liouville theorem follows
directly from the identity:
$$
\nabla\cdot(\nabla g(\vec{r}) \wedge \nabla h(\vec{r}))=0\; ,
$$
which tells us that the velocity field in Eq.~(\ref{a}) is
divergenceless.

As a physical motivation for Eq.~(\ref{a}), Nambu has shown that the
Euler equations for the angular momentum of a rigid body can be put
into that form  if the dynamical variables are taken to be the
components of the angular  momentum vector $\vec L = (L_x, L_y, L_z)$,
and $g$ and $h$ are taken to be,  respectively, the total kinetic
energy and the square of the angular momentum.

Moreover, he noticed that the evolution equation for a function $f$ on
${\Bbb R}^3$ induced by the equation of motion (\ref{a}) can be cast
into  the form:
\begin{equation}\label{c}
\frac{df}{dt}=\frac{\partial(f,g,h)}{\partial(x,y,z)}\; ,
\end{equation}
where the right-hand side is the Jacobian of $(f,g,h)$ with respect to
$(x,y,z)$. This expression was easily generalized to $n$
functions on ${\Bbb R}^n$. The Jacobian can be interpreted as a kind
of  generalized Poisson bracket: it is skew-symmetric with respect to
$f$, $g$ and $h$; it is a derivation of the algebra of smooth functions
on  ${\Bbb R}^3$, i.e., the Leibniz rule is verified in each argument.
Hence there is a complete analogy with the Poisson bracket
formulation of  Hamilton equations except, at first sight, for the
equivalent of Jacobi  identity which seems to be lacking. In fact, in
the usual Poisson formulation, the Jacobi identity is the
infinitesimal form of Poisson theorem which  states that the bracket
of two integrals of motion is also an integral of  motion. If we want
a similar theorem for Nambu Mechanics there must be an  infinitesimal
form of it which will provide a generalization of Jacobi
identity. Denote by $\{f,g,h\}$ the Jacobian appearing in
Eq.~(\ref{c}). Let $\phi_t\colon\vec{r}\mapsto\phi_t(\vec{r})$ be the
flow for Eq.~(\ref{a}).  Then a generalization of Poisson theorem
would imply that $\phi_t$  is a ``canonical transformation'' for the
generalized bracket:
$$
\{f_1\circ\phi_t,f_2\circ\phi_t,f_3\circ\phi_t\}=\{f_1,f_2,f_3\}
\circ\phi_t\; .
$$
Differentiation of this equality with respect to $t$ yields the
desired  generalization of Jacobi identity:
\begin{eqnarray*}
\{\{g,h,f_1\},f_2,f_3\}+\{f_1,\{g,h,f_2\},f_3\}+
\{f_1,f_2,\{g,h,f_3\}\}\\ =\{g,h,\{f_1,f_2,f_3\}\}\; , \quad \forall
g,h,f_1,f_2,f_3\in C^\infty({\Bbb R}^3).
\end{eqnarray*}
This identity and its generalization to ${\Bbb R}^{n}$, called
Fundamental  Identity (FI), was introduced by Flato, Fr\o nsdal
\cite{FF} and Takhtajan  \cite{Ta}  as a consistency condition for
Nambu Mechanics (this consistency  condition was also formulated in
\cite{SV}) and allows a generalized  Poisson theorem: the generalized
bracket of $n$ integrals of motion is an  integral of motion. It turns
out that the Jacobian on ${\Bbb R}^{n}$  satisfies the FI.

Since the publication of Nambu's paper in 1973, different aspects of
this  new geometrical structure have been studied by several
authors. In \cite{BF}, it is shown that Nambu Mechanics on ${\Bbb
R}^n$ can be viewed, through Dirac's constraints theory,  as an
embedding into a  singular Hamiltonian system on ${\Bbb R}^{2n}$. An
invariant geometrical  formulation of Nambu Mechanics has recently
been given in \cite{Ta}  leading to the notion of Nambu-Poisson
manifolds.  Several physical systems have been formulated within the
Nambu framework: in \cite{Ch}, it is shown,  among others, that the
$SU(n)$-isotropic harmonic  oscillator and the $SO(4)$-Kepler systems
admit a Nambu-Poisson structure.  Other examples are discussed in \cite{Ta}.
\subsection{An Overlook of Zariski Quantization}
Nambu also discussed the quantization of this new structure. This
turns out to be a non-straightforward task \cite{BF}, \cite{Ta} and
the usual  approaches to quantization failed to give an appropriate
solution.  See Sect.~2.1 for further details.

The aim of this paper is to present a solution for the quantization of
Nambu-Poisson structures. This solution is based on deformation quantization
and  involves arithmetic aspects in its construction related to
factorization of polynomials in several real variables. For that
reason,  the quantization scheme we shall present here is called
Zariski Quantization. We attack directly the question of deformation of
Nambu Mechanics as it stands by taking only into account the defining
relations (conditions a), b) and c) given below).
This problem of quantization of $n$-gebras (also closely related
to operads) is a very cute mathematical problem which we solve here
independently of any other scheme of quantization treated before.
It should also be mentioned that our quantization technique can be
applied to more general type of structures than Nambu-type structures.
We shall give here a brief overlook of this solution.

Consider the Nambu bracket on ${\Bbb R}^3$ given by the Jacobian:
\begin{equation}\label{aaa}
\{f_1,f_2,f_3\} = \sum_{\sigma\in S_3}\epsilon(\sigma)  \frac{\partial
f_1}{\partial x_{\sigma_1}}\frac{\partial f_2} {\partial x_{\sigma_2}}
\frac{\partial f_3}{\partial x_{\sigma_3}}\; ,
\end{equation}
where  $S_3$ is the permutation group of $\{1,2,3\}$ and
$\epsilon(\sigma)$  is the sign of the permutation~$\sigma$.
When one verifies that the Jacobian satisfies the FI,  all
one needs are some specific properties of the  pointwise product of
functions appearing in the right-hand side of Eq. (\ref{aaa}). Namely,
it is Abelian, associative, distributive  (with respect to addition)
and satisfies the Leibniz rule. The idea here is to look for a deformation
of the usual product which enjoys the previously stated properties and
to define a deformed Nambu bracket by replacing the usual product by
the deformed product. Denote by $\times$ such a deformed product.
Then the deformed bracket:
\begin{equation}\label{bbb}
[f_1,f_2,f_3] \equiv \sum_{\sigma\in S_3}\epsilon(\sigma)
\frac{\partial f_1}{\partial x_{\sigma_1}}\times\frac{\partial f_2}
{\partial x_{\sigma_2}}\times \frac{\partial f_3}{\partial
x_{\sigma_3}}\; ,
\end{equation}
will define a deformation of the Jacobian function expressed
by (\ref{aaa}).

In this desired context, the whole problem of quantizing
Nambu-Poisson structures
reduces to the construction of the deformed product $\times$. Some trivial
deformations of the usual product provide such deformed products, but
these are not interesting. Also one has to bear in mind a theorem by Gelfand
which states that an Abelian involutive Banach algebra $\cal B$ is
isomorphic to an
algebra of continuous functions on the spectrum (maximal ideals) of $\cal B$,
endowed with the pointwise product. Hence we cannot expect to find a
non-trivial deformation of the usual product on a dense subspace of
$C^0({\Bbb R}^{n})$ with all the desired properties. At best we would deform
the spectrum. Moreover, Abelian algebra deformations of Abelian algebras
are classified by the Harrison cohomology and it turns out that the
second Harrison cohomology space is trivial for an algebra of
polynomials \cite{GS1}. Hence it is not possible to find a non-trivial
Abelian algebra deformation (in the sense of Gerstenhaber \cite{GS1})
of the algebra of polynomials  on ${\Bbb R}^{n}$.

We shall see in Sect.~3.1 what are the difficulties met when one
tries to construct a deformed Abelian associative algebra consisting of
functions on ${\Bbb R}^{3}$. It is possible to construct an Abelian
associative deformation of the usual pointwise product on the space of
real polynomials on ${\Bbb R}^{3}$ of the following form:
\begin{equation}\label{ccc}
f\times_\beta g = T(\beta(f)\otimes \beta(g))\; ,
\end{equation}
where $\beta$ maps a real polynomial on ${\Bbb R}^3$ to the symmetric
algebra constructed over the polynomials on ${\Bbb R}^3$. $T$ is an
``evaluation map'' which allows to go back to (deformed)
polynomials. It replaces the (symmetric) tensor product $\otimes$ by a
symmetrized form of a ``partial'' Moyal product on ${\Bbb R}^3$
(Moyal product on a hyperplane in ${\Bbb R}^3$ with deformation
parameter $\hbar$).  The extension of the map $\beta$ to
deformed polynomials by requiring that it annihilates
(non-zero) powers of $\hbar$, will give rise to an Abelian deformation
of the usual product ($T$ restores a $\hbar$-dependence).
In general
(\ref{ccc}) does not define an associative product and we look for a
$\beta$ which makes the product $\times_\beta$ associative. Consider a real
(normalized) polynomial $P$ on ${\Bbb R}^3$: it can be uniquely
factored into irreducible factors $P= P_1 \cdots P_n$.  Define
$\alpha$ on the space of real (normalized) polynomials by: $\alpha(P)=
P_1\otimes \cdots \otimes P_n$.  With the choice $\beta=\alpha$ in (\ref{ccc}),
it can be easily shown that the product $\times_\alpha$ is associative.
But the map $\alpha$ is not a linear map, hence the product $\times_\alpha$
is not distributive and Leibniz rule is not verified. Note also that,
already at the product level (multiplicative semi-group of polynomials),
the obtained deformation is not of the type considered by Gerstenhaber
because the choice $\alpha(\hbar)=0$ does not allow base field extension
from $\Bbb R$ to  ${\Bbb R}[\hbar]$. The usual cohomological treatment
of deformations in the sense of Gerstenhaber is therefore not applicable here.

These difficulties are related to the fact that, from the physical point of
view, the dynamical variables with respect to which the Nambu bracket is
expressed do not necessarily represent point-particles (see the example
for Euler equations mentioned in Sect.~1.1).  As a matter of fact, the
point-particle interpretation in Hamiltonian Mechanics is based on the
following feature: one can construct dynamical systems with
phase-space of arbitrarily (even) dimension by composing systems with
phase-spaces of smaller dimensions.  Remember that ${\Bbb R}^{2n}$
endowed with its canonical Poisson bracket is nothing but the direct
sum of 2-dimensional spaces (${\Bbb R}^{2}$) endowed with their
canonical Poisson brackets. In this situation it is possible to
interpret a system of $n$ free particles as $n$ systems of one free
particle. Such a situation no longer prevails in Nambu Mechanics. The
FI imposes strong constraints on Nambu-Poisson structures and the
linear superposition
of two Nambu-Poisson structures does not define in general a
Nambu-Poisson structure
(see \cite{Ta}). In that sense, it seems hopeless to have some notion
of point-particles in Nambu Mechanics and this fact suggests that
quantization here will have more to do with a field-like approach than
with a quantum-mechanical one, and we shall have to quantize the
observables (functions) rather than the dynamical variables
themselves.

However a quantum-mechanical approach is possible
when the system under consideration deals with dynamical variables
for which a point-particle interpretation is lacking, i.e., without
position-momentum interpretation (e.g. the case of angular momentum).
Here the absence of linear superposition is natural since not physically
needed. One should then replace the Moyal product in the evaluation map
by an invariant (in general, covariant) star-product on the dual
of a Lie algebra $\frak g$. Also the map $\beta$ in (\ref{ccc}) is here 
linear and performs a complete factorization of monomials in the generators
(coordinates on ${\frak g}^\ast$) by:
$$\beta(L_1^{i_1}\cdots L_n^{i_n})
= L_1^{{i_1}\atop\otimes}\otimes\cdots\otimes L_n^{{i_n}\atop\otimes}\;,$$
where $L_1,\ldots,L_n$ are coordinates on ${\frak g}^\ast\sim {\Bbb R}^n$.
By imposing that the  map $\beta$ vanishes on the non-zero powers of $\hbar$,
the product $\times_\beta$ so obtained is associative and distributive and
provides an Abelian algebra deformation of the algebra
of polynomials on ${\frak g}^\ast$ endowed with the usual product.
Notice that in general the product $\times_\beta$ is not trivial.
The deformed Nambu bracket constructed with a non-trivial product
$\times_\beta$ will define a deformation of the Nambu-Poisson structure 
on ${\Bbb R}^n$. Hence in such a case there is no necessity for
a field-like quantization, we can quantize the dynamical variables
$L_1,\ldots,L_n$, and remain in a quantum-mechanical context, however 
not a canonical quantization.

When $\frak g$ is the Heisenberg algebra ${\frak h}_n$ with generators
$1,p_1,\ldots,p_n,q_1,\ldots,q_n$ the invariant star-product is the Moyal
product on ${\Bbb R}^{2n}$ and it turns out that the corresponding
product $\times_\beta$ is nothing but the usual product, i.e. no deformation
is obtained. Here  one cannot conciliate particle-interpretation
with quantization on the space of polynomials  and
one has to adopt a field-like point of view.

In relation with what has been said above about field-like
quantization for Nambu Mechanics and in order to get around Gelfand theorem
and cohomological difficulties,
we are led to consider an algebra ${\cal A}_0$ (a kind of Bosonic Fock
space) on which is defined the classical Nambu-Poisson structure:
quantization is interpreted as a (generalized) deformation ${\cal A}_\hbar$
of the algebra ${\cal A}_0$. More precisely, let ${\cal N}$ be an Abelian
associative algebra with product $(f,g)\mapsto f\cdot g$; the
algebraic structure of Nambu Mechanics is given by a trilinear map on
${\cal N}$ taking values in ${\cal N}$, $[\cdot,\cdot,\cdot]\colon
(f,g,h)\mapsto [f,g,h]\in {\cal N}$ such that $\forall
f_0,f_1,f_2,f_3,f_4,f_5\in{\cal N}$:
\begin{itemize}
\item[a)] $[f_1,f_2,f_3] =
\epsilon(\sigma)[f_{\sigma_1},f_{\sigma_2},f_{\sigma_3}]\;,
\quad \sigma\in S_3;$
\item[b)] $[f_0\cdot f_1,f_2,f_3]=f_0\cdot[f_1,f_2,f_3]+[f_0,f_2,f_3]\cdot
f_1\;;$
\item[c)] $[f_1,f_2,[f_3,f_4,f_5]]$

$\qquad\qquad =[[f_1,f_2,f_3],f_4,f_5]+
[f_3,[f_1,f_2,f_4],f_5] +[f_3,f_4,[f_1,f_2,f_5]]\;.$
\end{itemize}
This is the setting for classical Nambu Mechanics where the
algebra ${\cal N}$ is the algebra of smooth functions on ${\Bbb R}^{3}$
with the pointwise product, and the bracket is the Jacobian.
Since we are looking for a field-like quantization, the classical
Nambu Mechanics (and hence the Nambu bracket~(\ref{aaa})) will be
defined on a kind of Fock space algebra ${\cal A}_0$ with product $\bullet$, 
described in Sect.~3.3.
The map $\alpha$ is extended to ${\cal A}_0$ by linearity (with respect
to the addition in ${\cal A}_0$) and the classical evaluation map
defined above will take values in ${\cal A}_0$ and will simply replace the
symmetric tensor product by the usual product and the tensor sum
by the addition in ${\cal A}_0$.

Then quantization will consist in ``deforming" the algebra
$({\cal A}_0,\bullet)$ to an Abelian associative
algebra $({\cal A}_\hbar,\bullet_\hbar)$, by requiring that $\alpha$
annihilates $\hbar$ and by using the evaluation map which replaces the
symmetric tensor product by a symmetrized product given by a star-product.
The quantum Nambu bracket $[\cdot,\cdot,\cdot]_{\bullet_\hbar}$
will be given by expression~(\ref{bbb}) where the $\times$-product
is replaced by the $\bullet_\hbar$-product and where the derivatives are
defined on ${\cal A}_\hbar$. This extension will permit the FI and
the Leibniz rule (with respect to the bracket) to be satisfied. Hence
this deformed bracket on the algebra ${\cal A}_\hbar$ will define a
quantization of the classical Nambu-Poisson structure on ${\cal A}_0$. By the
same procedure, one gets immediately generalizations to ${\Bbb R}^n$,
$n\geq 2$.

The paper is organized as follows. Here below we review briefly Nambu-Poisson
manifolds. In Sect.~2 we discuss the problems encountered in quantization of
Nambu Mechanics and recall the deformation quantization approach. Section~3 is
devoted to the construction of a solution for the quantization of Nambu-Poisson
structures on ${\Bbb R}^n$, $n\geq2$, by introducing the Zariski quantization
scheme. The paper is concluded by several remarks about possible extensions of
this work and related mathematical problems.
\subsection{Nambu-Poisson Manifolds}
Let us first review some basic notions on Nambu-Poisson manifolds (the
reader is referred to \cite{Ta} for further details).  Let $M$ be a
$m$-dimensional $C^\infty$-manifold.  Denote by $A$ the algebra of
smooth real-valued functions on $M$.  $S_n$ stands for the group of
permutations of the set $\{1,\ldots,n\}$. We shall denote by
$\epsilon(\sigma)$ the sign of the permutation $\sigma\in S_n$.
\begin{definition} A Nambu bracket of order $n$ ($2\leq n\leq m$) on $M$
is defined by a $n$-linear map on $A$ taking values in $A$:
$$
\{\cdot,\ldots,\cdot\}\colon A^n\rightarrow A\; ,
$$
such that the following statements are satisfied $\forall
f_0,\ldots,f_{2n-1}\in A$:
\begin{itemize}
\item[a)] Skew-symmetry
$$
\{f_1,\ldots,f_n\} = \epsilon(\sigma)
\{f_{\sigma_1},\ldots,f_{\sigma_n}\}\; , \quad \forall \sigma\in S_n;
$$
\item[b)] Leibniz rule
\begin{equation}\label{h}
\{f_0f_1,f_2,\ldots,f_n\} =
f_0\{f_1,f_2,\ldots,f_n\}+\{f_0,f_2,\ldots,f_n\}f_1\; ;
\end{equation}
\item[c)] Fundamental Identity
\begin{eqnarray}\label{i}
&&\{f_1,\ldots,f_{n-1},\{f_{n},\ldots,f_{2n-1},\}\}\nonumber\\ &&\quad
=\{\{f_1,\ldots,f_{n-1},f_n\},f_{n+1},\ldots,f_{2n-1}\}\nonumber\\
&&\qquad
+\{f_n,\{f_1,\ldots,f_{n-1},f_{n+1}\},f_{n+2},\ldots,f_{2n-1}\}
\nonumber\\ &&\qquad +\cdots+\{f_{n},f_{n+1},\ldots,f_{2n-2},
\{f_1,\ldots,f_{n-1},f_{2n-1}\}\}\; .
\end{eqnarray}
\end{itemize}
\end{definition}
Properties a) and b) imply that there exists a $n$-vector field $\eta$
 on $M$ such that:
\begin{equation}\label{aa}
\{f_1,\ldots,f_n\}=\eta(df_1,\ldots,df_n)\; , \quad \forall
f_1,\ldots,f_n\in A.
\end{equation}
Of course the FI imposes constraints on $\eta$, analyzed in \cite{Ta}.
 A $n$-vector field on $M$ is called a Nambu tensor, if its associated
 Nambu bracket defined by Eq.~(\ref{aa}) satisfies the FI.
\begin{definition} A Nambu-Poisson manifold $(M,\eta)$ is a manifold $M$ on
 which is defined a Nambu tensor $\eta$.  Then $M$ is said to be
endowed with a Nambu-Poisson structure.
\end{definition}
The dynamics associated with a Nambu bracket on $M$ is specified by
$n-1$ Hamiltonians $H_1,\ldots,H_{n-1}\in A$ and the time evolution of
$f\in A$ is given by:
\begin{equation}\label{j}
\frac{df}{dt}=\{H_1,\ldots,H_{n-1},f\}\; .
\end{equation}
Suppose that the flow $\phi_t$ associated with Eq.~(\ref{j}) exists
 and let $U_t$ be the one-parameter group acting on $A$ by $f\mapsto
 U_t(f)=f\circ\phi_t$. It follows from the FI that:
\begin{theorem}
The one-parameter group $U_t$ is an automorphism of the algebra $A$
for the Nambu bracket.
\end{theorem}
\begin{definition}
$f\in A$ is called an integral of motion for the system defined by
 Eq.~(\ref{j}) if it satisfies $\{H_1,\ldots,H_{n-1},f\}=0$.
\end{definition}
It follows from the FI that a Poisson-like theorem exists for
Nambu-Poisson manifolds:
\begin{theorem}
The Nambu bracket of $n$ integrals of motion is also an integral of
motion.
\end{theorem}

For the case $n=2$, the FI is Jacobi identity and one recovers the
 usual definition of Poisson manifold.  On ${\Bbb R}^2$, the canonical
 Poisson bracket of two functions ${\cal P}(f,g)$ is simply their
 Jacobian and Nambu defined his bracket on ${\Bbb R}^n$ as a Jacobian
 of $n$ functions $f_1,\ldots,f_n\in C^\infty({\Bbb R}^n)$ of $n$
 variables $x_1,\ldots,x_n$:
$$
\{f_1,\ldots,f_n\} = \sum_{\sigma\in S_n}\epsilon(\sigma)
\frac{\partial f_1}{\partial x_{\sigma_1}}\cdots\frac{\partial f_n}
{\partial x_{\sigma_n}}\; ,
$$
which gives the canonical Nambu bracket of order $n$ on ${\Bbb R}^n$.
 Other examples of Nambu-Poisson structures have been found \cite{CT}.
 One of them is a generalization of linear Poisson structures and is
 given by the following Nambu bracket of order $n$ on ${\Bbb
 R}^{n+1}$:
$$
\{f_1,\ldots,f_n\} =\sum_{\sigma\in S_{n+1}}\epsilon(\sigma)
\frac{\partial f_1}{\partial x_{\sigma_1}}\cdots\frac{\partial f_{n}}
{\partial x_{\sigma_{n}}}x_{\sigma_{n+1}}\; .
$$
In general any manifold endowed with a Nambu-Poisson structure of
order~$n$ is locally foliated by Nambu-Poisson manifolds of dimension~$n$
endowed with the canonical Nambu-Poisson structure \cite{gau}. In particular,
it is shown in \cite{gau} that any Nambu tensor is decomposable
(this fact, conjectured in \cite{Ta}, was eventually discovered
to be a consequence of an old result \cite{W} reproduced in
a textbook by Schouten \cite{Scho} Chap. II Sects. 4 and 6, formula (6.7)).
\section{The Quantization Problem}
\subsection{Difficulties with Usual Quantizations}
In his 1973 paper Nambu has also studied the quantization of his
generalized mechanics. He was looking for an operator representation
of a trilinear bracket which is skew-symmetric and satisfies Leibniz
rule (several combinations of conditions weaker than the preceding
were discussed as well).  The main difficulty encountered was to
conciliate skew-symmetry and Leibniz rule at the same time. It is
interesting to note that Nambu suggested the use of non-associative
algebras in order to overcome the problems appearing with operatorial
techniques.

Other aspects of operatorial quantization of Nambu Mechanics were
discussed in \cite{BF}, \cite{CT}, \cite{Ta}.  In \cite{BF},
is performed an embedding of ${\Bbb R}^3$ into ${\Bbb R}^6$ and
the original Nambu Mechanics \cite{Na} is formulated in terms
of usual Hamiltonian flow with constraints. Under star-quantization
with constraints, one gets the quantization of Nambu Mechanics. This explains
the question of Nambu, namely: Why is it that classical Mechanics
can be ``generalized'' while Quantum Mechanics is ``so unique'' and is of
Heisenberg type? However this embedding is not canonical. In addition
this approach did not take into account the FI which was introduced much later.

In \cite{Ta}, a representation of the ($n=3$ case) Nambu-Heisenberg
commutation relations:
$$
[A_1,A_2,A_3]\equiv \sum_{\sigma\in S_3}
 \epsilon(\sigma)A_{\sigma_1}A_{\sigma_2}A_{\sigma_3}=cI\; ,
$$
where $c$ is a constant and $I$ is the unit operator, was constructed.
 The operators $A_1$, $A_2$, $A_3$ act on a space of states
 parametrized by a ring of algebraic integers ${\Bbb Z}[\rho]$ in the
 quadratic number field ${\Bbb Q}[\rho]$ (where
 $1+\rho+\rho^2=0$). The cases $n=5$ and $n=7$ are studied in
 \cite{CT}.

A possible alternative to quantize Nambu bracket by deformation
quantization \cite{BFFLSI}, \cite{BFFLSII} was discussed in \cite{Ta}
(see Sect.~2.2 for a brief review on star-products).  If one looks at
the canonical Nambu bracket on ${\Bbb R}^3$ as a trilinear
differential operator $D$ on $A=C^\infty({\Bbb R}^3)$, then one can
define a $\hbar$-deformed trilinear product on $A$ by:
\begin{equation}\label{2b}
(f_1,f_2,f_3)_\hbar=\exp (\hbar D)(f_1,f_2,f_3)\; ,\quad
f_1,f_2,f_3\in A.
\end{equation}
The ``deformed bracket'' associated with the product (\ref{2b}) would
naturally be defined by:
\begin{equation}\label{2c}
[f_1,f_2,f_3]_\hbar=\frac{1}{3!}\sum_{\sigma\in S_3}
\epsilon(\sigma)(f_{\sigma_1},f_{\sigma_2},f_{\sigma_3})_\hbar\; ,
\end{equation}
leading to a deformation of Nambu bracket. But (\ref{2c}) is not a
deformation of a Nambu-Poisson structure: {\it it does not satisfy the
FI}.  Furthermore, it is not clear what kind of associativity
conditions one should impose on a trilinear product for the Leibniz
rule to be valid. Anyhow, if $F$ is a nonlinear analytic function of
one variable, we know \cite{Gu} that there is no deformation of the
Nambu bracket satisfying the FI of the form:
$$
(f_1,f_2,f_3)_\hbar=F (\hbar D)(f_1,f_2,f_3)\; ,\quad f_1,f_2,f_3\in
A.
$$
Note that the previous negative result does not mean that there is no
 (differentiable) deformation of Nambu-Poisson structures since
 general deformations of the form:
$$
[f_1,f_2,f_3]_\hbar=\{f_1,f_2,f_3\}+\sum_{r\geq1}\hbar^r
D_r(f_1,f_2,f_3)\; ,
$$
where the $D_r$'s are trilinear differential operators on $A$, have to
 be considered, but it shows that deformation quantization will not
 provide a straightforward solution to the quantization problem of
 Nambu-Poisson structures. Nevertheless we shall present a solution in
 Sect.~3.3 that relies heavily on deformation quantization.

Another possible avenue for the quantization problem is to apply
Feynman Path Integral techniques. A canonical formalism and an action
principle have been defined for Nambu Mechanics permitting the
definition of an action functional \cite{Ta}. Within this formalism,
it would be possible to formally define the path integral for Nambu
Mechanics, but this approach is essentially equivalent to usual
deformation quantization since the Feynman Path Integral is given by
the star-exponential (see the end of Sect.~2.2).
\subsection{Deformation Quantization}
For completeness we give here a brief review on deformation
quantization and star-products; a full treatment can be found in
\cite{BFFLSI}, \cite{BFFLSII} and a recent review in \cite{FS}.  Let
$M$ be a Poisson manifold.  We denote by $A$ the algebra of
$C^\infty$-functions on $M$ and by ${\cal P}(f,g)$ the Poisson bracket
of $f,g\in A$. Let $A[[\nu]]$ be the space of formal power series in
the parameter $\nu$ with coefficients in $A$.  A star-product
$\ast_\nu$ on $M$ is an associative (generally non-abelian)
deformation of the usual product of the algebra $A$, and is defined as
follows:
\begin{definition}
A star-product on $M$ is a bilinear map $(f,g)\mapsto f\ast_\nu g$
from $A\times A$ to $A[[\nu]]$, taking the form:
$$
f\ast_\nu g = \sum_{r\geq0}\nu^r C_r(f,g)\; , \quad \forall f,g\in A,
$$
where $C_0(f,g)=fg$, $f,g\in A$, and $C_r\colon A\times A \rightarrow
A$ ($r\geq 1$) are bidifferential operators (bipseudodifferential
operators can sometimes be considered) on $A$ satisfying:
\begin{itemize}
\item[a)] $C_r(f,c)=C_r(c,f)=0$, $r\geq 1$, $c\in\Bbb R$, $f\in A$;
\item[b)] $C_1(f,g)-C_1(g,f)=2{\cal P}(f,g)$, $f,g\in A$;
\item[c)] $\displaystyle \sum_{\scriptstyle r+s=t\atop r,s\geq0}
 C_r(C_s(f,g),h) =\sum_{\scriptstyle r+s=t\atop r,s\geq0}
 C_r(f,C_s(g,h))$, $\forall t\geq0$, $f,g,h\in A$.
\end{itemize}
\end{definition}
By linearity, $\ast_\nu$ is extended to $A[[\nu]] \times A[[\nu]]$.
 Condition~a) ensures that $c\ast_\nu f = f\ast_\nu c=cf$, $c\in \Bbb
 R$ (and may be omitted, in which case an equivalent star-product
 will verify it).
 Condition~c) is equivalent to the associativity equation $(f\ast_\nu
 g)\ast_\nu h= f\ast_\nu(g\ast_\nu h)$.  Condition~b) implies that the
 star-bracket
$$
[f,g]_{\ast_\nu}\equiv (f\ast_\nu g -g \ast_\nu f)/2\nu\; ,
$$
is a deformation of the Lie-Poisson algebra on $M$. Hence a
star-product on $M$ deforms at once the two classical structures on
$A$, i.e. the Abelian associative algebra for the pointwise product of
functions and the Lie algebra structure given by the Poisson
bracket. This leads to:
\begin{definition}
A deformation quantization of the Poisson manifold $(M,{\cal P})$ is a
star-product on $M$.
\end{definition}
\begin{definition} Two star-products $\ast$ and $\ast^\prime$ are said to be
 equivalent if there exists a map $T\colon A[[\nu]] \rightarrow
 A[[\nu]]$ having the form:
$$
T=\sum_{r\geq0}\nu^r T_r\; ,
$$
where the $T_r$'s $(r\geq1)$ are differential operators vanishing on
 constants and $T_0=Id$, such that
$$
Tf\ast Tg = T(f\ast^\prime g)\; ,\quad f,g\in A[[\nu]].
$$
A star-product which is equivalent to the pointwise product of
 functions is said to be trivial.
\end{definition}
For physical applications, the deformation parameter $\nu$ is taken to
be $i\hbar/2$. On ${\Bbb R}^{2n}$ the basic example of star-product is
the Moyal product defined by:
\begin{equation}\label{2i}
f\ast^{}_{M}g=\exp\left(\frac{i\hbar}{2}{\cal P}\right)(f,g)\; .
\end{equation}
It corresponds to the Weyl (totally symmetric) ordering of operators
in Quantum Mechanics. On ${\Bbb R}^{2n}$ endowed with its canonical
Poisson bracket, other orderings can be considered as well and they
correspond to star-products equivalent to the Moyal product. For
example, the normal star-product (which is the exponential of ``half
of the Poisson bracket'' in the variables $p\pm iq$) is equivalent
to Moyal product. {}From now on, we implicitly set $\nu=i\hbar/2$.

A given Hamiltonian $H\in A$ determines the time evolution of an
observable $f\in A$ by the Heisenberg equation:
\begin{equation}\label{2h}
\frac{df_t}{dt}=[H,f_t]_{\ast_\nu}\; .
\end{equation}
The one-parameter group of time evolution associated with
Eq.~(\ref{2h}) is given by the star-exponential defined by:
\begin{equation}\label{exp}
\exp_{\ast}\left(\frac{tH}{i\hbar}\right) \equiv \sum_{r\geq0}
\frac{1}{r!}  \left(\frac{t}{i\hbar}\right)^{r}(\ast H)^r\; ,
\end{equation}
where $(\ast H)^r=H\ast\cdots\ast H$ ($r$ factors).  Then the solution
to Eq.~(\ref{2h}) can be expressed as:
$$
f_t = \exp_{\ast}\left(\frac{tH}{i\hbar}\right)\ast f \ast
\exp_{\ast}\left(\frac{-tH}{i\hbar}\right)\; .
$$
In many examples, the star-exponential is convergent as a series in
the variable $t$ in some interval ($|t|< \pi$ for the harmonic
oscillator in the Moyal case) and converges as a distribution on $M$
for fixed $t$.  Then it makes sense to consider a Fourier-Dirichlet
expansion of the star-exponential:
\begin{equation}\label{2n}
\exp_{\ast}\left(\frac{tH}{i\hbar}\right)(x)= \int \exp(\lambda t/i
\hbar)d\mu(x;\lambda)\; ,\quad x\in M,
\end{equation}
the ``measure'' $\mu$ being interpreted as the Fourier transform (in
the distribution sense) of the star-exponential in the variable $t$.
Equation (\ref{2n}) permits to define \cite{BFFLSII} the spectrum of
the Hamiltonian $H$ as the support $\Lambda$ of the measure $\mu$. In
the discrete case where
$$
\exp_{\ast}\left(\frac{tH}{i\hbar}\right)(x)=\sum_{\lambda\in \Lambda}
\exp(\lambda t/i \hbar)\pi_\lambda(x)\; ,\quad x\in M,
$$
the functions $\pi_\lambda$ on $M$ are interpreted as eigenstates of
$H$ associated with the eigenvalues~$\lambda$, and satisfy
$$
H\ast\pi_\lambda =\pi_\lambda \ast H=\lambda \pi_\lambda\; , \quad
\pi_\lambda \ast \pi_{\lambda'}=\delta_{\lambda\lambda'}\; , \quad
\sum_{\lambda\in \Lambda} \pi_\lambda = 1\; .
$$
In the Moyal case, the Feynman Path Integral can be expressed
\cite{Pa} as the Fourier transform over momentum space of the
star-exponential.  In field theory, where the normal star-product is
relevant, the Feynman Path Integral is given (up to a multiplicative
factor) \cite{Di} by the star-exponential.

{}From the preceding, it should be clear that deformation quantization
provides a completely autonomous quantization scheme of a classical
Hamiltonian system and we shall use it for the quantization of
Nambu-Poisson structures.
\section{A New Quantization Scheme:
Zariski Quantization} We saw in Sect.~2.1 that a direct application of
deformation quantization to Nambu-Poisson structures is not
possible. Instead of looking at the deformed Nambu bracket as some
skew-symmetrized form of a $n$-linear product, we deform directly the
Nambu bracket. Then it turns out that a solution to the quantization
problem can be constructed in this way,
based on the following simple remark: the Jacobian
of $n$ functions on ${\Bbb R}^n$ is a Nambu bracket because the usual
product of functions is Abelian, associative, distributive and
respects the Leibniz rule.  If we replace the usual product in the
Jacobian by any product having the preceding properties, we get a
``modified Jacobian'' which is still a Nambu bracket.  That is to say,
the ``modified Jacobian'' is skew-symmetric, it satisfies the Leibniz
rule with respect to the new product and the FI is verified. Now if we
suppose that the new product is a deformation of the usual product,
then the ``modified Jacobian'' will be a deformation of the Nambu
bracket providing a deformation quantization of the Nambu-Poisson
structure.
\subsection{Quantization of Nambu-Poisson Structure of Order $3$:\penalty-10000
The Setting} This section is devoted to preliminaries needed for
the construction of an Abelian
associative deformed product on ${\Bbb R}^3$. The generalization to
${\Bbb R}^n$ will be discussed later.

First we shall make some general comments on possible candidates that
 one can consider for an Abelian deformed product. Even though ${\Bbb
 R}^3$ is not a symplectic manifold, we can define a ``partial'' Moyal
 product between functions in $A=C^\infty({\Bbb R}^3)$. Denote by
 $(x_1,x_2,x_3)$ the coordinates in ${\Bbb R}^3$. Let ${\cal P}_{12}$
 be the Poisson bracket with respect to the variables $(x_1,x_2)$,
 i.e. for $f,g\in A$, it is defined by ${\cal
 P}_{12}(f,g)=\frac{\partial f}{\partial x_1}\frac {\partial
 g}{\partial x_2}- \frac{\partial f}{\partial x_2}\frac{\partial g}
 {\partial x_1}$.  Then denote by $\ast_{12}$ the Moyal product
 constructed with ${\cal P}_{12}$ and with deformation parameter
 $\hbar$, that is:
$$
f\ast_{12} g=\sum_{r\geq 0} \frac{\hbar^r}{r!} {\cal P}_{12}^r(f,g)\;,
\quad f,g\in A.
$$
Then $A[[\hbar]]$ endowed with the product $\ast_{12}$ is a
non-abelian associative deformation of $A$ endowed with the usual
product. If, in order to get an Abelian algebra, one simply applies
the ``Jordan trick'' to the non-abelian algebra
$(A[[\hbar]],\ast_{12})$ by defining a product by $f\times
g=\frac{1}{2}(f\ast_{12}g+g\ast_{12}f)$, one will get a
non-associative algebra. Here associativity is lacking, because the
product $\times$ does not make a complete symmetrization with respect
to $(f_1,f_2,f_3)$ in the expression $(f_1\times f_2)\times f_3$.

Somehow a kind of symmetrization, not necessarily with respect to the
factors appearing in the product, is needed for associativity and the
product we are looking for should share some features of the tensor
product of particle-states in the Bosonic Fock space as is done in
second quantization.  It suggests to look at a map sending $f\in A$ to
the symmetric tensor algebra $\hbox{\rm Symm}(A)$ of $A$ and then go
back to $A[[\hbar]]$ by an ``evaluation map'' which replaces the
symmetric tensor product in $\hbox{\rm Symm}(A)$ by a completely
symmetrized form of the Moyal product $\ast_{12}$.

Let us make precise the previous remark. Start with any map:
$$ \beta\colon A \rightarrow \hbox{\rm Symm}(A)\; ,$$
such that $\beta(1)=I$ and extend it to the map from $A[[\hbar]]$ into
$\hbox{\rm Symm}(A)$ (denoted by the same symbol $\beta$)
by requiring that it vanishes on the non-zero
powers of $\hbar$. Define the evaluation map $T\colon\hbox{\rm Symm}(A)
\rightarrow A[[\hbar]]$ as a canonical linear map whose restriction
on $A^{n\atop\otimes}$ is given by:
\begin{equation}\label{2q}
f_1\otimes\cdots\otimes f_n \mapsto \frac{1}{n!}\sum_{\sigma\in S_n}
f_{\sigma_1}\ast_{12}\cdots\ast_{12} f_{\sigma_n}\; ,
\end{equation}
where $\otimes$ stands for the {\it symmetric\/} tensor product.
Then we define a map $\times_{\beta}\colon A[[\hbar]] \times A[[\hbar]]
\rightarrow A[[\hbar]]$ --- the $\beta$-product --- by the following formula:
\begin{equation}\label{2Q}
f\times_\beta g=T(\beta(f)\otimes\beta(g))\; ,\quad f,g \in A[[\hbar]].
\end{equation}
It is clear that the $\beta$-product is always Abelian. However, for a general
map $\beta$, it is neither associative, nor distributive, nor a deformation of
the usual pointwise product on A, or has $1$ as unit element. Thus
associativity $(f \times_{\beta}g) \times_{\beta} h=f \times_{\beta}
(g\times_{\beta} h)$ of the $\beta$-product reads
$$
T(\beta(T(\beta(f) \otimes \beta(g)))\otimes \beta(h))=
T(\beta(f) \otimes \beta(T(\beta(g) \otimes \beta(h))))\;,
\quad \forall f,g,h \in A[[\hbar]],
$$
and it is an equation for the map $\beta$.
Before giving a non-trivial example
of the associative $\beta$-product which is a deformation of the usual
product (the expression ``deformation'' being given a broad sense as
explained in the proof of Theorem~4 below, i.e., a $\hbar$-dependent product
whose limit at $\hbar=0$ is the initial product),
we summarize simple basic facts regarding this construction in the
following theorem.
\begin{theorem}
\begin{itemize}
\item[i)] The standard unit $1$ is the unit element of the
$\beta$-product: $f \times_{\beta}1$ $=f$, $\forall f \in A$,
if and only if $T \circ \beta=id_{A}$.
\item[ii)] If, in addition to i), $\beta\colon A \rightarrow
\hbox{\rm Symm}(A)$ is an algebra
homomorphism, then the $\beta$-product on $A$ coincides with the usual
pointwise product.
\item[iii)]
If the $\beta$-product is a deformation of the usual product, then the
associativity condition reduces to
$$T
(\beta(fg) \otimes \beta(h))=T(\beta(f) \otimes\beta(gh))\;,
\quad \forall f,g,h \in A.$$
\item[iv)]
If, in addition to i), the $\beta$-product is an associative deformation of
the usual product, then it coincides with the usual product.
\item[v)]
If $\beta$ is an algebra homomorphism and the $\beta$-product is a
deformation of the usual product, then the $\beta$-product is associative.
\end{itemize}
\end{theorem}
\begin{proof} Part i) is obvious, since it is equivalent to
$$f \times_{\beta}1=T(\beta(f))=f\; ,\quad  \forall f \in A.$$
For the part ii), we have
$$f \times_{\beta} g=T(\beta(f) \otimes \beta(g))=T(\beta(fg))=fg\; ,
\quad \forall f,g \in A.$$
To prove iii), simply note that if the $\beta$-product is a deformation of the
usual product, then $\beta(f \times_{\beta} g)=\beta(fg)$ ($fg$ stands for the
usual product), and the equation follows.

Part iv) follows from part iii) by setting $h=1$ in the associativity
condition, then we get:
$$
f \times_{\beta} g=T(\beta(fg))=fg\; ,\quad \forall f,g \in A.$$
Finally, part v) follows from part iii) and associativity of
the $\otimes$-product in  $\hbox{\rm Symm}(A)$, since
\begin{eqnarray*}
T(\beta(fg) \otimes \beta(h))&=&T((\beta(f) \otimes \beta(g))
\otimes \beta(h))=T(\beta(f) \otimes (\beta(g) \otimes \beta(h)))\\
&=&T(\beta(f) \otimes \beta(gh))\;.
\end{eqnarray*}
\end{proof}

We shall give an example of such map $\beta$ for which condition v) of the
Theorem~3 is satisfied, so that it gives an Abelian associative deformation
of the usual product. For that purpose, we need to restrict $A$
to the algebra $N$ of
polynomials on ${\Bbb R}^3$, which will allow a more refined
decomposition in the
symmetric algebra, thus avoiding the triviality of the product.
In fact, we shall factorize polynomials on ${\Bbb R}^3$
into irreducible factors $P=P_1 \cdots P_n$ and send them to elements
of the form $P_1 \otimes \cdots \otimes P_n$ in the symmetric algebra.
This will give the desired Abelian associative deformation of the
usual product.
\begin{remark}
The standard embedding of the polynomial algebra into its
symmetric algebra by elements of degree 1 (i.e.~without any decomposition at
all) gives rise to a non-associative product because of the incompatibility
between the usual product and the Moyal product: associativity would require
that $(PQ)*R+R*(PQ)=P*(QR)+(QR)*P$, which fails in general.
\end{remark}
\begin{remark}
Another extreme case is when every polynomial is embedded
into the symmetric algebra via complete symmetrization
(i.e.~by replacing every
monomial by the corresponding $\otimes$-monomials in the symmetric algebra).
In this example $\beta$-product again gives the usual product. Indeed
the corresponding map $\beta$ is obviously a homomorphism and, according to
Theorem~3, part ii), one needs to verify that $T(\beta(P))=P$ for all
polynomials $P$, that is to say, $T(Q_1\otimes\cdots \otimes Q_n)=
Q_1\cdots Q_n$, where $Q_i$ stands for $x_1$, $x_2$ or $x_3$.
This fact represents a well-known property of the Moyal
quantization, and its proof is left to the reader.
\end{remark}

The choice of $\beta$ we shall present makes a non-trivial compromise
between commutativity and associativity though at this stage it lacks
the property of being distributive (i.e. we deform at first only the
semi-group structure); the construction, in the original phase-space setting,
is nevertheless interesting in itself.
This $\beta$-product is constructed as follows.
Let $N={\Bbb R}[x_1,x_2,x_3]$ be the algebra of polynomials in the
variables $x_1$, $x_2$, $x_3$ with real coefficients and let
$${\cal S}(N)=\bigoplus^{\infty}_{n=1}N^{n\atop\otimes}\; ,$$
be its symmetric tensor algebra without scalars. Next, for any $P \in N$
define its {\it maximal monomial\/} to be a monomial of the highest total
degree in $P$, maximal with respect to the lexicographical ordering induced
by $(x_1, x_2, x_3)$. We call $P \in N$ a {\it normalized\/} polynomial, if its
maximal monomial has coefficient $1$. Since the product of normalized
polynomials is again normalized, normalized polynomials form a semi-group
that we shall denote by $N_1$.
We should also include $0$ as a normalized polynomial, so that $0 \in
N_1$.

Also consider the algebra $N[\hbar]$ (polynomials in $\hbar$ with coefficients
in $N$) and call $P \in N[\hbar]$ a normalized polynomial if the
coefficient of its lowest degree term in $\hbar$ is normalized in $N$.
All normalized
polynomials in $N[\hbar]$ form a semi-group $N^{\hbar}_1$ (under the usual
product).

Every polynomial in $N_1$ can be uniquely factored into a product of
irreducible normalized polynomials:
$$P=P_1 \cdots P_n\; .$$
Note that this factorization, as well as the set of all irreducible
polynomials, depend on the choice of the ground field (in our case
${\Bbb R}$, the field of real numbers).
Since we are dealing with polynomials in several variables, even over
the field of complex numbers irreducible polynomials need
not to be linear. In fact, the set of all irreducible polynomials in $n$
variables over a  field $k$ plays a fundamental role in algebraic geometry
over $k$: it defines the so-called Zariski topology in the space $k^n$
(and in the corresponding projective space as well). This is why we call the
concrete realization of the $\beta$-product, based on the factorization of
polynomials, Zariski quantization.

We  define a  map $\tilde{\alpha}\colon N_1 \rightarrow {\cal S}(N)$ by:
$$
\tilde\alpha (P)= P_1\otimes\cdots\otimes P_n\; , \quad P\in N_1.
$$
Denote by $\pi\colon  N^{\hbar}_1 \rightarrow N_1$ the homomorphism
which attaches to a polynomial in $N^{\hbar}_1$ its coefficient of
degree $0$ in $\hbar$; it is always an element in $N_1$ (and may be zero
as well). This ``projection onto the classical part'' allows to extend
$\tilde{\alpha}$ to the  homomorphism
$$
\alpha=\tilde{\alpha} \circ \pi\colon N^{\hbar}_1 \rightarrow{\cal S}(N)\; ,
$$
which takes into account only the classical part of the polynomial in
$N^{\hbar}_1$.  Finally, denote the restriction of the evaluation map $T$ from
$\hbox{\rm Symm}(A)$ to ${\cal S}(N)$ by the same symbol
$T\colon {\cal S}(N) \rightarrow N[\hbar]$. Then
specializing our general construction of $\beta$-products to the case
$\beta=\alpha$ we get the map
$\times_\alpha\colon N^{\hbar}_1 \times N^{\hbar}_1 \rightarrow N[\hbar]$,
given by the following formula:
$$P \times_\alpha Q=T(\alpha(P) \otimes \alpha(Q))\; ,\quad  \forall P,Q
\in N^{\hbar}_1.$$
\begin{theorem}
The map $\times_\alpha$ defines an Abelian associative product on
$N^{\hbar}_1$ which is a deformation of the usual product on $N_1$.
\end{theorem}
\begin{proof}
First, the classical part of $P \times_\alpha Q$ is equal to
$\pi(P)\pi(Q) \in N_1$ (it may be zero as well), since the classical part
of the Moyal product is the usual product and $\pi$ is a homomorphism. This
shows that indeed the map $\times_\alpha$
maps $N^{\hbar}_1 \times N^{\hbar}_1$
into $N^{\hbar}_1$. In particular, if $P,Q \in N_1$, then
$$P \times_\alpha Q|_{\hbar=0}=PQ\; ,$$
so that $\times_\alpha$ is some deformation of the usual product.
By this we mean nothing more than the above formula; due to the
projection onto the classical part and the decomposition into
irreducible factors, what we get is more general
than a deformation in the sense of Gerstenhaber; in particular
``Gerstenhaber" deformations are defined on the base field
${\Bbb R}[[\hbar]]$ while here (at least in the present construction)
we do not have $\hbar$-linearity.
Second, $\alpha$ is a homomorphism, so that associativity follows
from Theorem~3, part v).
\end{proof}
\begin{remark}
Note that in the definition of the evaluation map $T$ the Moyal
product $*_{12}$ can be replaced by any star-product on ${\Bbb R}^3$ without
affecting the associativity and deformation properties of the product
$\times_\alpha$. In particular, one also has products $\times_\alpha^{(ij)}$
constructed from partial Moyal products on $(ij)$-planes in ${\Bbb R}^3$.
It is easy to show that the totally  symmetrized product:
$(f,g)\mapsto \frac{1}{3}(f\times_\alpha^{(12)} g +
f\times_\alpha^{(23)} g +
f\times_\alpha^{(31)} g)$, is an Abelian, associative deformation
of the usual product.
\end{remark}
\begin{remark}
Note that $1$ is not a unit element for the product
$\times_\alpha$. Indeed, in general it is not true that $P \times_\alpha 1=P$,
$\forall P \in N^{\hbar}_1$. However, it is true when $P$ is either an
irreducible polynomial, or reduces completely into a product of
linear factors.
\end{remark}

The space $N_1^\hbar$ endowed with the product $\times_\alpha$ is then an
Abelian semi-group. The following example shows that
($N_1^\hbar,\times_\alpha$)
cannot be extended to an algebra in $N[\hbar]$. Consider the
polynomials $P=x_1^2+\epsilon^2 x_2^2$, $\epsilon\in \Bbb R$, and
$Q=x_2^2$.  $P$ is irreducible, then $\alpha(P)=x_1^2+\epsilon^2
x_2^2$ (considered as an element of $N^{1\atop\otimes}$), while
$\alpha(Q)=x_2\otimes x_2\in N^{2\atop\otimes}$. One has (for notation
simplicity, we write here $\ast$ instead of $\ast_{12}$)
\begin{eqnarray*}
&&P\times_\alpha Q\\ &&\quad = T((x_1^2+\epsilon^2 x_2^2)\otimes x_2\otimes
x_2)\\ &&\quad =\frac{1}{3}\big[(x_1^2+\epsilon^2 x_2^2)\ast x_2\ast
x_2+x_2\ast (x_1^2+\epsilon^2 x_2^2)\ast x_2 + x_2\ast x_2 \ast
(x_1^2+\epsilon^2 x_2^2)\big]\\ &&\quad =(x_1^2+\epsilon^2 x_2^2)x_2^2
+\frac{2}{3}\hbar^2\; .\\
\end{eqnarray*}
It is easy to verify that $x_1^2\times_\alpha x_2^2 = x_1^2 x_2^2$ and
$x_2^2\times_\alpha x_2^2=x_2^4$, so we have
$(x_1^2+\epsilon^2 x_2^2)\times_\alpha x_2^2 \neq x_1^2
\times_\alpha x_2^2 + \epsilon^2
(x_2^2\times_\alpha x_2^2)$.  Hence $\times_\alpha$ is
{\it not a distributive\/} product
with respect to the addition in $N[\hbar]$.  Moreover the preceding
example shows that: $ \lim_{\epsilon\rightarrow 0}((x_1^2+\epsilon^2
x_2^2)\times_\alpha x_2^2) \neq x_1^2\times_\alpha x_2^2$,
i.e. $\times_\alpha$ is {\it not a continuous\/} product.

These special  aspects of $\times_\alpha$ imply the following: if we replace
the usual product in the canonical Nambu bracket of order 3 by the
product $\times_\alpha$ in order to get a deformed Nambu bracket:
$$
[f,g,h]_\hbar \equiv \sum_{\sigma\in S_3} \epsilon(\sigma)
\frac{\partial f}{\partial x_{\sigma_1}}\times_\alpha \frac{\partial g}
{\partial
x_{\sigma_2}}\times_\alpha \frac{\partial g}{\partial x_{\sigma_3}}\; ,
$$
we will not get a deformation of the Nambu-Poisson structure. It can
be easily verified that the Leibniz rule (with respect to $\times_\alpha$) and
the FI are not satisfied. At this point, these facts should not be
too surprising: as mentioned in Sect.~1.2, we know that we cannot
expect to find a non-trivial deformation of the
usual product on $N$ with all the nice properties.

To summarize, we have some space $N_1$ with the usual product, and a deformed
product on $N_1^\hbar$.  Along the lines of what is done for
topological quantum groups \cite{BFGP} and in second quantization,
let us look at ``functions'' on $N_1$ (e.g. formal
series). Intuitively we get a deformed coproduct and the dual of this
space of ``functions'' (polynomials on polynomials) will then have a
product and a deformed product, both of which will be distributive
with respect to the vector space addition.  Now the product of
polynomials is again a polynomial. So in fact we are getting some
deformed product on an algebra generated by the
polynomials. We shall make this heuristic view precise in the next
section.
\subsection{Zariski Product}
The product $\times_\alpha$ on $N_1^\hbar$ defined in Sect.~3.1 is
Abelian and associative, but is not distributive with respect to the addition
in $N[\hbar]$. Hence $(N_1^\hbar,\times_\alpha)$ is only a semi-group.
We shall extend the product $\times_\alpha$ to an algebra
${\cal Z}_\hbar$ and get an Abelian algebra deformation of an Abelian algebra
${\cal Z}_0$ generated by the irreducible polynomials in $N_1$. The
algebra ${\cal Z}_0$ is actually a kind of Fock space constructed from
the irreducible polynomials considered as building blocks.

Let $N^{irr}_1 \subset N_1$ be the set of real irreducible normalized
polynomials. Let ${\cal Z}_0$ be a real vector space having a basis indexed
by products of elements of $N^{irr}_1$, we denote the basis by
$\{Z_{u_1\cdots u_m}\}$, where $u_1,\ldots, u_m\in N^{irr}_1$, and $m \geq 1$.
The vector space ${\cal Z}_0$ is made into an algebra by defining
a product $\bullet^z\colon{\cal Z}_0\times{\cal Z}_0\rightarrow{\cal Z}_0$ by:
$$
Z_{u_1\cdots u_m}\bullet^z Z_{v_1\cdots v_n}
=Z_{u_1\cdots u_m v_1\cdots v_n}\; ,
\quad\forall u_1,\ldots, u_m, v_1,\ldots v_n \in N_1,\forall m,n\geq1.
$$
${\cal Z}_0$ endowed with the product $\bullet^z$ is the
free Abelian algebra generated by the set of irreducible polynomials or
equivalently the algebra of the semi-group $N_1$. Note that the addition
in ${\cal Z}_0$ is {\it not} related to the addition in $N$, i.e.
$Z_{u+v}\neq Z_{u}+Z_{v}$.

Every $u\in N$ can be uniquely factored as follows: $u=cu_1\cdots u_m$,
where $c\in\Bbb R$ and $u_1,\ldots, u_m\in N_1$, and we shall sometimes write
$Z_u$ for $c Z_{u_1\cdots u_m}$. This provides a multiplicative (but non
additive) injection of $N$ into the algebra ${\cal Z}_0$.

Let ${\cal Z}_\hbar={\cal Z}_0[\hbar]$ be the vector space of polynomials
in $\hbar$ with coefficients in ${\cal Z}_0$. Let the map $\zeta\colon
N_1^\hbar
\rightarrow {\cal Z}_\hbar$ be the injection of $N_1^\hbar$ into
${\cal Z}_\hbar$ defined by:
\begin{equation}\label{ppp}
\zeta(\sum_{r\geq0}\hbar^r u_r)=\sum_{r\geq0}\hbar^r Z_{u_r}\; ,
\quad \forall u_0\in N_1, u_i\in N, i\geq 1.
\end{equation}
Using the injection $\zeta$ we can extend the product $\times_\alpha$
on $N_1^\hbar$ to ${\cal Z}_\hbar$ by first defining the product
on the basis elements:
\begin{equation}\label{productz}
Z_{u_1\cdots u_m}\bullet^z_\hbar Z_{v_1\cdots v_n}
=\zeta((u_1\cdots u_m)\times_\alpha( v_1\cdots v_n))\; ,
\end{equation}
$\forall u_1,\ldots, u_m, v_1,\ldots v_n \in N_1$,
$\forall m,n\geq1$, and then extend it to all of ${\cal Z}_\hbar$ by requiring
that the product $\bullet^z_\hbar$ annihilates the non-zero powers of $\hbar$:
$$
(\sum_{r\geq0}\hbar^r A_r)\bullet^z_\hbar
(\sum_{s\geq0}\hbar^s B_s)=A_0\bullet^z_\hbar B_0\; ,
\quad \forall A_r, B_s \in {\cal Z}_0, r,s \geq 0.
$$
\begin{theorem}
The vector space ${\cal Z}_\hbar$ endowed with the product $\bullet^z_\hbar$
is an Abelian algebra which is some deformation of the Abelian algebra
$({\cal Z}_0,\bullet^z)$.
\end{theorem}
\begin{proof}
By definition the product $\bullet^z_\hbar$ is distributive and Abelian.
The associativity of $\bullet^z_\hbar$ follows directly from the associativity
for the product $\times_\alpha$. For $\hbar=0$, the product $\times_\alpha$
is the usual product, and Eq.~ (\ref{productz}) becomes, with
$u=u_1\cdots u_m$ and $v=v_1\cdots v_n$:
$$
Z_{u}\bullet^z_\hbar Z_{v}|_{\hbar=0}
=\zeta(uv)=Z_{uv}=
Z_{u}\bullet^z Z_{v}\; ,
$$
showing that the product $\bullet^z_\hbar$ is some deformation of
the product $\bullet^z$.
\end{proof}

The next step would be to define derivatives $\delta_i$,
$1\leq i\leq 3$, on ${\cal Z}_0$ and then
extend them to ${\cal Z}_\hbar$. This would allow to define first
the classical Nambu bracket on ${\cal Z}_0$, and the quantum one on
${\cal Z}_\hbar$. The ``trivial'' definition $\delta_i Z_u=Z_{\partial_i u}$,
$\forall u\in N$, where $\partial_i$ is the usual derivative
with respect to $x^i$, does not satisfy the Leibniz rule
(except on the diagonal, a remark relevant for the deformed exponential
(\ref{3a})) because
of the different nature of the addition in $N$ and in ${\cal Z}_0$.

Unfortunately, what seems to be another  very natural
definition of derivative on ${\cal Z}_0$ does not satisfy the
Frobenius property (commutativity of the derivatives in several variables,
a property that was trivially satisfied by the previous ``trivial'' definition
for which Leibniz rule did not hold).
These derivatives would be linear maps
$\delta_i\colon {\cal Z}_0 \rightarrow{\cal Z}_0$,
$1\leq i\leq 3$, defined as follows. For $u\in N^{irr}_1$, we let
$\delta_i Z_u = Z_{\partial_i u}$, where $\partial_i$ denotes the usual
partial derivative of $u$ with respect to $x^i$. The action of $\delta_i$
on a general basis element $Z_v$, $v\in N_1$, is given by postulating Leibniz
rule on the product of irreducible polynomials $v=v_1v_2\cdots v_m$:
$$
\delta_i\ Z_{v_1v_2\cdots v_m}=
Z_{(\partial_iv_1)v_2\cdots v_m}+
Z_{v_1(\partial_iv_2)\cdots v_m}+\cdots +
Z_{v_1v_2\cdots (\partial_iv_m)}\; .
$$
Obviously, the maps $\delta_i$
are derivations on the algebra ${\cal Z}_0$, but one can easily show
that they are not commuting maps,
i.e. $\delta_i\delta_j\neq \delta_j\delta_i$, $i\neq j$. This comes from
the fact that when one takes the derivatives of an irreducible polynomial $u$,
the polynomials $\partial_i u$, $1\leq i\leq 3$, do not necessarily
factorize out into the same number of factors.
An example is given in ${\Bbb R}^2$ by
$u = (x^3 + x^2y + 4xy^2 + 5y^3 + 5xy + {17 \over 2}y^2 + 4y) \in N_1^{irr}$.
A consequence of this fact is the following: If one defines the classical
Nambu bracket on ${\cal Z}_0$ by replacing, in the Jacobian,
the usual product by $\bullet^z$ and the usual partial derivatives
by the maps $\delta_i$, this new bracket will not satisfy the
FI. There will be anomalies in the FI (even at this classical, or
``prequantized" level) due to terms
which can not cancel out each other because
the Frobenius property is not satisfied on ${\cal Z}_0$.
In order to have a family of commuting derivations
which can naturally be related to the usual derivatives of a polynomial,
we need to extend the algebra on which will be defined
the classical Nambu bracket. This algebra will consist of Taylor series
in the variables $(y^1,y^2,y^3)$ of the translated polynomials $u(x+y)$.
One can look at this algebra as a jet space over the polynomials
and it will be constructed in the next section.

Nonetheless the algebra ${\cal Z}_\hbar$ with product $\bullet^z_\hbar$
provides an Abelian deformation of the algebra ${\cal Z}_0$ and
this also is interesting {\it per se\/} because it gives
an example of a non-trivial Abelian deformation, however generalized
and therefore not necessarily classified by
the Harrison cohomology (defined on the sub-complex of the
Hochschild complex consisting of symmetric cochains \cite{GS1}, \cite{GS2}).
\subsection{Quantization of Nambu-Poisson Structure of order $3$:\penalty-10000
A Solution}
Let us construct the space ${\cal A}_0$ on which will be
defined the classical Nambu-Poisson structure. On this space we will have
an injection of the semi-group $N_1$ (normalized polynomials in
the variables $(x^1,x^2,x^3)$) which will allow a natural definition
of the derivative of an element of ${\cal A}_0$. We shall consider a
space of ``Taylor series'' in the variables $(y^1,y^2,y^3)$ of translated
polynomials $x\mapsto u(x+y)$ with coefficients in the algebra ${\cal Z}_0$
introduced in Sect.~3.2.

Denote by ${\cal E}={\cal Z}_0[y^1,y^2,y^3]$, the algebra of polynomials
in the variables $(y^1,y^2,y^3)$ with coefficients in ${\cal Z}_0$.
Instead of the usual Taylor series
$$ u(x+y) = u(x) + \sum_{i}y^i {\partial_i u}(x) + {1\over 2}\sum_{i,j}
y^i y^j {\partial_{ij} u}(x) + \cdots\;, $$
which we multiply by $(uv)(x+y)=u(x+y)v(x+y)$
we look at ``Taylor series'' in ${\cal E}$, for $u \in N_1$:
\begin{equation}\label{yyy}
  J(Z_u)=Z_u + \sum_{i}y^i Z_{\partial_i u} + {1\over 2}\sum_{i,j}
y^i y^j Z_{\partial_{ij} u} + \cdots
= \sum_{n} {1 \over {n!}} (\sum_{i}y^i\partial_i)^n(Z_u)\;,
\end{equation}
where $\partial_{i} u$, $\partial_{ij} u$, etc. are the usual derivatives
of $u\in N_1\subset N$ with respect to the variables $x^i$, $x^i$ and
$x^j$, etc., ${\partial_i} Z_u \equiv Z_{\partial_i u}$ and, since in
general the derivatives of $u\in N_1$ are in $N$,
one has to factor out the appropriate constants in
$Z_{\partial_i u}$, $Z_{\partial_{ij} u}$, etc. (i.e. $Z_{\lambda u}\equiv
\lambda Z_u$, $u\in N_1$, $\lambda\in \Bbb R$). $J$ defines an additive
map from ${\cal Z}_0$ to ${\cal E}$
(to say that $J$ is multiplicative is tantamount to the Leibniz property).

Let ${\cal A}_0$ be the sub-algebra of ${\cal E}$ generated by elements
of the form (\ref{yyy}). We shall denote by $\bullet$ the product in
${\cal A}_0$ which is naturally induced by the product in ${\cal E}$.
In order to define the (classical) Nambu-Poisson structure on ${\cal A}_0$,
we need to make precise what is meant by the derivative of an element
of ${\cal A}_0$. Remember that the derivative ${\partial_i}u(x+y)$ is
again a Taylor series of the form ${\partial_i}u(x)
+ \sum_j y^j{\partial_{ij}}u(x)+ \cdots$.
We shall define thus the derivative
$\Delta_a$, $1\leq a\leq 3$, of an element
of the form (\ref{yyy}) by the natural extension to ${\cal A}_0$ of the
previous ``trivial'' definition, i.e.,
\begin{equation}\label{yyya}
\Delta_a(J(Z_u))=J(Z_{\partial_a u})= Z_{\partial_a u} +
\sum_{i}y^i Z_{\partial_{ai} u} + {1\over 2}\sum_{i,j}
y^i y^j Z_{\partial_{aij} u} + \cdots\;,
\end{equation}
for $u\in N_1$, $1\leq a\leq 3$. One can look at  definition (\ref{yyya})
of $\Delta_a$ as the restriction,
to the subset of elements of the form $J(Z_u)$, of the formal derivative
with respect to $y^a$ in the ring ${\cal E}={\cal Z}_0[y^1,y^2,y^3]$.
Since $\Delta_a(J(Z_u))=J(Z_{\partial_a u})$,
we have $\Delta_a({\cal A}_0)={\cal A}_0$ and we get a family of maps
$\Delta_a\colon {\cal A}_0\rightarrow{\cal A}_0$, $1\leq a \leq 3$,
restriction to ${\cal A}_0$ of the derivations with respect to $y^a$,
$1\leq a \leq 3$, in ${\cal E}$. We can summarize the properties of $\Delta_a$
in  the:
\begin{lemma}
The maps $\Delta_a\colon {\cal A}_0\rightarrow{\cal A}_0$, $1\leq a\leq 3$,
defined by Eq.~(\ref{yyya}) constitute a family of commuting
derivations (satisfying the Leibniz rule) of the algebra ${\cal A}_0$.
\end{lemma}
\begin{proof}
Follows directly from the fact that the $\Delta_a$, $1\leq a\leq 3$, are
the restrictions to the sub-algebra ${\cal A}_0$ of the formal derivatives
in $y^a$ on the ring ${\cal E}={\cal Z}_0[y^1,y^2,y^3]$.
\end{proof}

The definition of derivatives on ${\cal A}_0$ leads
to the following natural definition of the classical Nambu bracket
on the Abelian algebra ${\cal A}_0$:
\begin{definition}
The classical Nambu bracket on ${\cal A}_0$ is the trilinear
map taking values in ${\cal A}_0$ given by:
\begin{equation}\label{xxx}
(A,B,C)\mapsto [A,B,C]_\bullet \equiv \sum_{\sigma\in S_3}
\epsilon(\sigma) \Delta_{\sigma_1}A\bullet \Delta_{\sigma_2}B\bullet
 \Delta_{\sigma_3}C\; ,
\quad \forall A,B,C\in {\cal A}_0.
\end{equation}
\end{definition}
\begin{theorem}
The classical Nambu bracket given in Def.~7 defines a Nambu-Poisson structure
on ${\cal A}_0$.
\end{theorem}
\begin{proof}
It follows trivially from the fact that $({\cal A}_0,\bullet)$ is an Abelian
algebra and from Lemma~1.
\end{proof}

Now that we have a classical Nambu-Poisson structure on ${\cal A}_0$, we shall
construct a quantum Nambu-Poisson structure by defining
some Abelian deformation
$({\cal A}_\hbar,\bullet_\hbar)$ of $({\cal A}_0,\bullet)$. The construction
is based on the map $\alpha$ introduced in Sect.~3.1 and we shall
extend the definition of the product $\bullet_\hbar^z$ defined in Sect.~3.2
to the present setting for the Nambu-Poisson structure on  ${\cal A}_0$.

Let ${\cal E}[\hbar]$ be the algebra of polynomials in $\hbar$
with coefficients in ${\cal E}$. We consider the subspace  ${\cal A}_\hbar$
of ${\cal E}[\hbar]$ consisting of series $\sum_{r\geq0}\hbar^r A_r$
for which the  coefficient $A_0$
is in ${\cal A}_0$. Then we define a map
$\bullet_\hbar\colon {\cal A}_\hbar\times {\cal A}_\hbar\rightarrow
{\cal E}[\hbar]$ by extending the product $\bullet_\hbar^z$
defined by (\ref{productz}) (it is sufficient to define it on ${\cal A}_0$
since $\bullet_\hbar^z$ annihilates the non-zero powers of $\hbar$):
\begin{equation}\label{product}
J(Z_u)\bullet_\hbar J(Z_v) = Z_u\bullet_\hbar^z Z_v
+ \sum_{i} y^i ( Z_{\partial_i u}\bullet_\hbar^z Z_v +  Z_u\bullet_\hbar^z
Z_{\partial_i v})+\cdots\;,\quad\forall u,v\in N_1.
\end{equation}
Actually $\bullet_\hbar$ defines a product on ${\cal A}_\hbar$ and we have:
\begin{theorem}
The vector space ${\cal A}_\hbar$ endowed with the product
$\bullet_\hbar$ is an  Abelian algebra which is some Abelian
deformation of the Abelian algebra $({\cal A}_0,\bullet)$.
\end{theorem}
\begin{proof}
For $A=\sum_{r\geq0}\hbar^r A_r$ and $B=\sum_{s\geq0}\hbar^s B_s$ in
${\cal A}_\hbar$, we have $A\bullet_\hbar B =A_0\bullet_\hbar B_0$ and
the coefficient of $\hbar^0$ of the latter is $A_0\bullet B_0$ which is in
${\cal A}_0$ since $A_0,B_0\in {\cal A}_0$.
This shows that $\bullet_\hbar$ is actually a product
on ${\cal A}_\hbar$. By definition this product is Abelian. Hence
$({\cal A}_\hbar,\bullet_\hbar)$ is an Abelian algebra.

It is clear from the preceding that for $\hbar=0$,
we have $A\bullet_\hbar B|_{\hbar=0} = A_0\bullet B_0$, which
shows that the product $\bullet_\hbar$ is some deformation
of the product $\bullet$.
\end{proof}

The derivatives $\Delta_a$, $0\leq a\leq 3$, are naturally extended
to ${\cal A}_\hbar$. Every element $A\in {\cal A}_\hbar$
can be written as $A=\sum_I y^I A_I$, where $I=(i_1,\ldots,i_n)$ is a
multi-index and $A_I\in {\cal Z}_\hbar$. Then the product $A\bullet_\hbar B$,
$A,B\in {\cal A}_\hbar$, reads:
$$
A\bullet_\hbar B = \sum_{I,J} y^I y^J A_I\bullet^z_\hbar B_J\;.
$$
Since $({\cal Z}_\hbar,\bullet^z_\hbar)$ is an Abelian algebra and
the derivative $\Delta_a$ acts as a formal derivative with respect
to $y^a$ on the product $A\bullet_\hbar B$, the usual properties
(linearity, Leibniz, Frobenius) of a derivative are still satisfied
on ${\cal A}_\hbar$. So we  can now define the quantum Nambu
bracket on ${\cal A}_\hbar$.
\begin{definition}
The quantum Nambu bracket on ${\cal A}_\hbar$ is the trilinear
map taking values in ${\cal A}_\hbar$ defined by:
\begin{equation}\label{xxxq}
(A,B,C)\mapsto [A,B,C]_{\bullet_\hbar} \equiv \sum_{\sigma\in S_3}
\epsilon(\sigma) \Delta_{\sigma_1}A\bullet_\hbar \Delta_{\sigma_2}B
\bullet_\hbar \Delta_{\sigma_3}C\; ,
\quad \forall A,B,C\in {\cal A}_\hbar.
\end{equation}
\end{definition}
\begin{theorem}
The quantum Nambu bracket endows ${\cal A}_\hbar$
with a Nambu-Poisson structure which is some deformation of the classical Nambu
structure on ${\cal A}_0$
\end{theorem}
\begin{proof}
The proof that the quantum Nambu bracket endows ${\cal A}_\hbar$
with a Nambu-Poisson structure is similar to the one of Theorem~6. That the
quantum Nambu bracket is some deformation of the classical Nambu
bracket follows from Theorem~7.
\end{proof}
\subsection{Generalizations}
What has been done in the previous two sections can be easily
generalized to ${\Bbb R}^n$, $n\geq 2$.  The only non-straightforward
modification to be done appears in the evaluation map (\ref{2q}). One
has to distinguish two cases: when $n$ is even and when $n$ is odd. If
$n=2p$, $p\geq 1$, then one replaces the partial Moyal product in
(\ref{2q}) by the usual Moyal product on ${\Bbb R}^{2p}$.  If
$n=2p+1$, $p\geq 1$, one uses the partial Moyal product
$\ast_{1\cdots2p}$ on the hyperplane defined by $x_{2p+1}=0$ (as for
the case $n=3$, other possibilities can be considered).  The other
definitions and properties are directly generalized to ${\Bbb R}^n$.
Note that the canonical Nambu-Poisson structure of order 2 on
${\Bbb R}^2$ is the usual Poisson structure; there our procedure gives a
quantization of the Poisson bracket ${\cal P}$ different from Moyal,
however not on $N[\hbar]$ but on ${\cal A}_\hbar$;
this quantization will in a sense be
somewhat like in field theory.  The same applies to ${\Bbb R}^{2n}$ by
starting with a sum of Poisson brackets on the various ${\Bbb R}^{2}$.

Our construction can be generalized to any orbit of the coadjoint
action of a Lie algebra on its dual (the case of ${\Bbb R}^3$ corresponds
to ${\frak{su}}(2)^\star$). In that case, instead of the Moyal product
appearing in the evaluation map, one can use a covariant star-product
on the orbit \cite{FS}.
\section{Concluding Remarks}
We have found a quantized version of Nambu Mechanics and we shall end this
 article with a few remarks concerning some related physical and mathematical
 points. We would like to stress that many features of the solution proposed
can be of direct relevance for other quantization problems.
\subsection{Sesqui-quantization}
One should notice that here we quantize a linear span of polynomials which
are in a way our ``fields''. In this scheme the irreducible polynomials play
 a very special r\^ole: they generate all the polynomials and are kind of
 building blocks in the quantum case. For example
on ${\Bbb R}^2$ the harmonic oscillator Hamiltonian
 $H=\frac{1}{2}(p^2+q^2)$ cannot be considered as the sum
of the two observables $p^2$ and $q^2$; it has to be considered as an
 irreducible  element of the algebra. The same thing is true for the
 anharmonic oscillator with Hamiltonian
 $\frac{1}{2}(p^2 + q^2 +\lambda q^4)$, $\lambda>0$, which is not
considered here as the sum of a free Hamiltonian with an interaction term.
 In usual Quantum Mechanics the Hermitian operator $H=P^2+Q^2$ is the sum
of two operators, but the physically  measurable quantities (spectrality)
 related to these operators seem to ignore that the Hamiltonian is the sum
of two observables. To make it precise, the spectrum of the harmonic
oscillator Hamiltonian is discrete, while $P^2$ and $Q^2$ have both
 continuous positive spectra; hence a priori there is no  way to relate these
 spectra. Before going further, let us mention that, as in the usual
deformation quantization case, we have a natural definition of
the spectrum of an observable in the Zariski Quantization.
Consider the polynomial $H\in N_1$ in the variables $p$ and $q$,
and map it to its Taylor series $J(Z_H)\in {\cal A}_0$ given by
(\ref{yyy}), and build the deformed exponential function:
\begin{equation}\label{3a}
\exp_{{\bullet_\hbar}}\left(\frac{tH}{i\hbar}\right)= \sum_{n\geq 0}
 \frac{1}{n!}\left(\frac{t}{i\hbar}
\right)^n J(Z_H){\bullet_\hbar}\cdots{\bullet_\hbar} J(Z_H)\; .
\end{equation}
In (\ref{3a}) let $y=0$, then:
$$
J(Z_H){\bullet_\hbar}\cdots{\bullet_\hbar} J(Z_H)|_{y=0}
= Z_H{\bullet_\hbar}^z\cdots{\bullet_\hbar}^z Z_H =\zeta(H\times_\alpha
\cdots \times_\alpha H) \in {\cal Z}_\hbar\; ,
$$
where $\zeta$ is defined in (\ref{ppp}).
As for the star-exponential (\ref{exp})
we define the spectrum of $H$ to be the support of the measure appearing
in the Fourier-Dirichlet expansion of (\ref{3a}) with $y=0$.
For $H$ irreducible, it is easy to see that
 $J(Z_H){\bullet_\hbar}\cdots{\bullet_\hbar} J(Z_H)|_{y=0}=
\zeta(H\ast\cdots\ast H)$,
where $\ast$ is the Moyal product on
 ${\Bbb R}^2$ with deformation parameter $\frac{1}{2}i\hbar$. In that case
we get the same spectrum as for the Moyal case. For completely reducible
 elements like $p^2$ and $q^2$, the exponential (\ref{3a}) reduces to the
usual exponential, in which case the spectrum is continuous.
So these three observables have the same spectra as in the usual case,
 and the Zariski Quantization scheme makes a distinction among them from
the very beginning. Somehow this new scheme is halfway between
first and second quantizations (hence the name ``sesqui-quantization''):
it is not quite a field theory (though a field-like formulation is possible)
but shares many features with it (Fock space, irreducible polynomials seen
 as ``1-particle'' states, etc.).
\subsection{Zariski Star-Products}
For the Poisson case, Zariski Quantization gives a quantization
which differs from the usual one in many respects. The most important one is
that the quantum Poisson bracket is {\it not\/} the skew-symmetrized form
of an associative product. But this quantized bracket can be seen as the
 ``classical'' part  of another quantum bracket coming from an
associative algebra. For ${\Bbb R}^{2n}$, consider the Zariski-Poisson
 bracket ${\cal P}_{{\bullet_\hbar}}$ built as indicated in Sect.~3.4;
in definition
(\ref{2i}) of the Moyal product replace  ${\cal P}$ by
${\cal P}_{{\bullet_\hbar}}$; we get the Zariski-Moyal product
 $f\bullet_M g =\exp(\nu {\cal P}_{{\bullet_\hbar}})$, where
$\nu$ is at first seen as a different parameter and later identified with
$\frac{1}{2}i\hbar$. Due to the properties of ${\bullet_\hbar}$, one gets
another
associative deformation of the usual product.
The corresponding deformed bracket will then start with the (Zariski)
quantum Poisson bracket and provides a  Lie algebra deformation.
Then a theory of ``star-products'' constructed with the product
${\bullet_\hbar}$ can be
developed in a  straightforward way.
\subsection{General Poisson Manifolds: An Overlook}
The quantization presented here was done in an algebraic setting, the product
${\bullet_\hbar}$ being  defined on the algebra ${\cal A}_\hbar$
constructed from polynomials on
${\Bbb R}^n$.
One can consider extensions to an algebraic variety $S$. It  should be
possible  to define a similar Abelian deformed product between polynomials
 on $S$ using an embedding of $S$ into ${\Bbb R}^n$ by polynomial
Dirac constraints \cite{Dir} that will induce on $S$ a Poisson structure.
 Furthermore we know from Nash \cite{Nas} that compact real analytic
 Riemannian manifolds can be analytically and isometrically embedded into some
${\Bbb R}^n$; the proof follows from his previous
 result on differentiable embeddings by showing that  there are
``arbitrarily close'' analytic and differentiable manifolds.
In this context, it is thus reasonable to expect that the procedure
 developed here can be extended (at least in the compact case) to arbitrary
 differentiable manifolds. Eventually, as for Nambu Mechanics,
 similar techniques may be applied to the quantization of not necessarily
regular Poisson structures on algebraic varieties, real analytic manifolds
 and differential manifolds.
\subsection{Cohomology}
{}From a mathematical point of view, it would be interesting to study general
 Abelian deformations of ${\cal Z}_0$ and ${\cal A}_0$ and look for
 associated cohomology complexes. A more detailed study of
 the kind of ``deformation" obtained here for these algebras, both as
 associative algebras and as Nambu bracket algebras, is certainly
 worthwhile. In view of Sect.~4.2, ``quantum''  cohomology  versions
 of the relevant cohomologies should also be of interest.
\begin{acknowledgement}
The authors wish to thank C. Fr\o nsdal, O. Mathieu, G. Pinczon,\penalty-10000
W.~Schmid, A.~Voronov, our colleagues at RIMS, and especially
Jacques~C.~H.~Simon for stimulating and very useful discussions.
The first three authors also thank RIMS and especially its
director H.~Araki for superb hospitality. M.F. thanks the Japanese Ministry
of Education, Science, Sports and Culture for financial support as
Invited Professor at RIMS.
\end{acknowledgement}

\end{document}